# Magnetic anisotropy related to hybridization between Fe 3*d* and As 4*p* orbitals in a bcc Fe-As thin film

Takahito Takeda[1†], Karumuri Sriharsha[1], Seiji Aota[1], Ryo Okano[1], Le Duc Anh[1,2],
Yukiharu Takeda[3], Akira Yasui[4], Miho Kitamura[5], Yuki K. Wakabayashi[6],
Atsushi Fujimori[7,8], Masaaki Tanaka[1,9,10], and Masaki Kobayashi[1,9*]

[1]*Department of Electrical Engineering and Information Systems, The University of Tokyo, 7-3-1 Hongo, Bunkyo-ku, Tokyo 113-8656, Japan*

[2]*Institute of Engineering Innovation. The University of Tokyo, 7-3-1 Hongo, Bunkyo-ku, Tokyo 113-0032, Japan*

[3]*Materials Sciences Research Center, Japan Atomic Energy Agency, Sayo-gun, Hyogo 679-5148, Japan*

[4]*Japan Synchrotron Radiation Research Institute, 1-1-1 Kouto, Sayo, Hyogo 679-5198, Japan*

[5]*Photon Factory, Institute of Materials Structure Science, High Energy Accelerator Research Organization (KEK), 1-1 Oho, Tsukuba 305-0801, Japan*

[6]*NTT Basic Research Laboratories, NTT Corporation, Atsugi, Kanagawa 243-0198, Japan*

[7]*Department of Physics, The University of Tokyo, 7-3-1 Hongo, Bunkyo-ku, Tokyo 113-0033, Japan*

[8]*Department of Physics and Center for Quantum Technology, National Tsing Hua University, Hsinchu 30013, Taiwan*

[9]*Center for Spintronics Research Network, The University of Tokyo, 7-3-1 Hongo, Bunkyo-ku, Tokyo 113-8656, Japan*

[10]*Institute for Nano Quantum Information Electronics, The University of Tokyo, 4-6-1, Komaba, Meguro-ku, Tokyo 153-8505 Japan*

Corresponding authors: [†]ttakeda@g.ecc.u-tokyo.ac.jp, [*]masaki.kobayashi@ee.t.u-tokyo.ac.jp

## ABSTRACT

The magnetic anisotropy (MA) of Fe-based ferromagnetic thin films has been extensively studied for device applications. The examined material is a new Fe-based ferromagnetic thin film, bcc $Fe_{1-x}As_x$ (Fe-As) with the in-plane MA (IMA) grown on a GaAs (111)B substrate. The magnetic properties of the Fe-As thin film have been investigated by X-ray magnetic circular dichroism (XMCD) and magnetic circular dichroism in hard X-ray photoemission spectroscopy (MCD-HAXPES) to elucidate the role of As ions in the IMA. The XMCD spectra at the Fe $L_{2,3}$ edge and MCD-HAXPES spectra of the Fe $2p$ core level exhibit ferromagnetic and metallic features like Fe metal. The XMCD at the As $L_{2,3}$ edge demonstrates that the As ions contribute to the ferromagnetism of bcc Fe-As through the hybridization between the Fe $3d$ and As $4p$ orbitals. The estimations of the magnetic moments of Fe using the XMCD sum rules have revealed that the orbital magnetic moment is isotropic and the magnetic dipole term is anisotropic. The anisotropy of the magnetic dipole term can be attributed to the anisotropic $p$-$d$ hybridization due to epitaxial strain, contributing to the IMA of bcc Fe-As. Our findings enlighten the mechanism of the MA of the non-magnetic ion-doped bcc Fe thin film, which can be applied to other magnetic $3d$ transition metal thin films doped with non-magnetic elements.

## 1. Introduction

The magnetic properties of ferromagnetic materials have been studied for their applications in various fields, such as data storage, magnetic sensors, and spintronics [1–3]. One of the fundamental properties of ferromagnetic materials is their magnetic anisotropy (MA), which is the tendency of a material to exhibit a preferred direction of magnetization. Understanding the origin of the MA is crucial for designing and optimizing materials for device applications. Fe metal and Fe compounds have attracted significant attention as ferromagnets for such applications. Because the bulk bcc Fe metal exhibits ferromagnetism with cubic MA [4], Fe thin films are dominated by shape anisotropy. Various approaches have been proposed to induce MA, particularly perpendicular magnetic anisotropy (PMA), in Fe compounds by alloying Fe with other elements (CoFeB/MgO [5], NdFeB [6,7], FePt alloy [8,9] and $Fe_xN$ [10,11]) or by making Fe ultrathin films (Fe 8 monolayer/GaAs [12] and Fe 0.7 nm/MgO [13,14]).

Apart from Fe, the PMA of other ferromagnetic materials has been studied so far, such as CoPt alloy [15,16], $Mn_{1-x}Ga_x$ [17,18], and $Cr_2Ge_2Te_6$ [19]. In order to elucidate the mechanism of the MA, X-ray magnetic circular dichroism (XMCD) is a powerful experimental method. By XMCD, one can investigate the magnetic properties of each element in ferromagnetic compounds [20–23]. A general explanation of the MA in Fe compounds such as FePt alloy [24,25] and Fe ultrathin films [14,26] is the changes in the spin-orbit interaction and/or orbital magnetic moment from more isotropic Fe metal. However, the mechanism of the MA in Fe compounds doped with non-magnetic elements, such as $Fe_xN$, is still under study [11,27]. Clarifying the mechanism of the MA in rare-earth/other-transition-metal/heavy-metal free Fe-based ferromagnets is essential for both understanding of the underlying physics in the MA and applications of these materials.

Recently, Aota *et al*. have successfully grown a bcc $Fe_{1-x}As_x$ (Fe-As) film on a GaAs (111)B substrate by molecular beam epitaxy (MBE) [28]. The bcc Fe-As film is a ferromagnetic metal with in-plane magnetic anisotropy (IMA) and high Curie temperature ($T_C$) above 400 K. Knowledge of the MA in bcc Fe-As develops the understanding of the role of doped non-magnetic ions in simple Fe alloys for the MA. In this study, we have enlightened the contribution of As ions to the MA in bcc Fe-As by XMCD and magnetic circular dichroism in hard X-ray photoemission spectroscopy (MCD-HAXPES). Here, we have performed MCD-HAXPES measurements to characterize the electronic structure and magnetic properties of Fe. The experimental findings reveal the relationship between the MA and alloying with As, which would contribute to understanding the MA of Fe metal induced by doping non-magnetic elements.

## 2. Experimental

An Fe-As thin film with a thickness of 14 nm was grown on a semi-insulating (SI) GaAs (111)B substrate using MBE. During the MBE growth, the surface conditions were monitored by reflection high energy electron diffraction (RHEED). The RHEED pattern of the Fe-As film in the [$\bar{1}10$] direction shown in Fig. 1(a) indicates that the Fe-As film was grown with a smooth surface. The ratio of Fe to As atoms was estimated to be ~3:1 by energy dispersive X-ray spectroscopy. The film was covered by an amorphous As capping layer to avoid surface oxidization. The sample structure was (from top to bottom) an As capping layer ~2 nm / Fe-As 14 nm / semi-insulating (SI)-GaAs (111)B substrate, as shown in the inset of Fig. 1(a). The epitaxial relationship between bcc Fe-As and GaAs is represented in Fig. 1(b). The microscopic crystal structure of the Fe-As film projected along the $[11\bar{2}]$ axis was observed by high angular annular dark field scanning transmission electron microscopy (HAADF-STEM) [see Fig. 1(c)], representing that the Fe-As film was epitaxially grown on the GaAs (111) substrate. The X-ray diffraction (XRD) $2\theta$-$\omega$ scan of the Fe-As film shown in Fig. 1(d) suggests that the crystal quality of the measured film is very high, similar to that in the previous study [28]. The lattice constant $c$ along the [111] axis was 0.501 nm estimated from the XRD measurement as shown in Fig. 1(d), and the in-plane lattice constant $a$ was the same as GaAs, that is, 0.399 nm estimated by the STEM image shown in Fig. 1(c). The values of $c$ and $a$ are longer and shorter than those of bulk bcc Fe, 0.498 nm and 0.405 nm, respectively, indicating that the Fe-As film on the GaAs (111) substrate was under compressive epitaxial strain. The Fe-As film in this paper is in the same category as the sample A in Ref [28]. X-ray absorption spectroscopy (XAS) and XMCD experiments were conducted at beamline BL23SU of SPring-8 [29]. The measurements were performed at a temperature of 10 K. The directions of the applied magnetic fields were parallel to the incident X-ray and perpendicular to the sample surface, which is defined as $\theta = 0°$, and tilted to $\theta = 60°$ as shown in Fig. 2(a). XAS and XMCD signals were measured in the total electron-yield mode. XMCD spectra were averaged as $\left((\mu_{+,+h} - \mu_{-,+h}) - (\mu_{+,-h} - \mu_{-,-h})\right)/2 \left(= \Delta\mu\right)$, and each XAS spectrum was obtained as $(\mu_{+,+h} + \mu_{-,+h})/2 \ (= \bar{\mu})$. Here, $\mu_+$ and $\mu_-$ denote the absorption of the X-rays with the polarization $\sigma_+$ (the photon helicity parallel to the spin polarization) and $\sigma_-$ (antiparallel to the spin polarization), and $+h$ ($-h$) means the magnetic field directions parallel (antiparallel) to the wave vector of the incident beam. For the XMCD measurements, the samples were etched in hydrochloric acid and rinsed in water just before loading them into the vacuum chambers to obtain clean surfaces. The MCD-HAXPES experiments were performed at beamline BL09XU of SPring-8 [30]. The angle between the hemispherical analyzer and the incident X-ray was 90°. The incident X-ray enters the sample surface with a grazing angle of 3°. The experimental setup is summarized in Fig. 2(b). The measurements were conducted at a temperature of 50 K.

The incident photon energy was fixed at 7.94 keV with a probing depth of several tens of nm. The Fermi level ($E_F$) position has been corrected using the Fermi edge of Au in electrical contact with the samples, and the binding energy ($E_B$) is measured relative to $E_F$. To magnetize the film along the in-plane direction, a permanent magnet approached the film at $T = 50$ K just before measurements. As for HAXPES, we measured an unetched sample since the large probing depth of HAXPES provides the bulk properties rather than the surface features.

## 3. Results and discussion

We have conducted SQUID measurements to elucidate the magnetic properties of the whole sample, as shown in Figs. 3(a) and 3(b). In Fig. 3(a), diamagnetic backgrounds from the GaAs substrate are subtracted by fitting the magnetization curves in the region between $\mu_0H = 2$ T and 4 T to a linear function, and the perpendicular magnetization curve is without the demagnetizing field correction. The saturation magnetization ($m_{sat,\text{SQUID}}$) of the film was estimated as 1.72 $\mu_B$/atom [see Figs. 3(a)]. The Fe-As thin film shows ferromagnetism up to 400 K with IMA [Fig. 3(b)]. The element-specific magnetic properties of the Fe-As thin film have been investigated by XMCD. The Fe $L_{2,3}$ XAS and XMCD spectra under $\mu_0H = 4$ T, as shown in Fig. 4(a), have smooth spectral line shapes without any multiplet structure and are similar to those of Fe metal [31], representing the metallic feature of the bcc Fe-As thin film. Note that the ferromagnetic component of Fe-As is saturated above $\mu_0H = 2$ T irrespective of the direction of $\mu_0H$, as shown in Fig. 3(d). The small but finite XMCD intensity of As at the $L_{2,3}$ absorption edge has been observed [see Fig. 4(b)]. The observation suggests that the As ions are magnetized in the bcc Fe-As thin film. The As $L_3$ XMCD spectrum shows the two finite peaks ($P_1$ and $P_2$), where the peaks of $P_1$ and $P_2$ in the As $L_3$ XMCD spectrum originate from the $2p \rightarrow 4s$ and $2p \rightarrow 4d$ excitations, respectively [32,33]. While the change of the orbital angular momentum in the $P_2$ process is the same as that of the Fe $L_3$ edge ($2p \rightarrow 3d$), where the difference of the orbital angular momentum is +1, the sign of XMCD at $P_2$ is opposite to the Fe $L_3$ XMCD sign. This result indicates that the magnetic moment of the As ions is antiparallel to that of the Fe ions. This antiferromagnetic coupling suggests that the electronic stricture and magnetic properties of Fe ions are modulated by the doped As ions and *vice versa* because As is a non-magnetic element.

We have conducted MCD-HAXPES to study the modification of the electronic structure of bcc Fe by As doping. The left panel of Fig. 5 shows the Fe $2p$ HAXPES and the MCD spectra of the Fe-As thin film. Here, the spectra of a 20-nm Fe metal film are also shown as a reference [34]. The spectral shapes of the MCD-HAXPES spectra have no distinct difference between the bcc Fe-As and 20-nm Fe metal film. The comparison suggests that the electronic structure of the ferromagnetic Fe component in Fe-As is similar to that of the 20-nm Fe metal film. Moreover, the peak position of Fe $2p_{3/2}$ MCD of Fe-As is ~200 meV lower than that of the Fe films, as shown in the right panel of Fig. 5. The similarity of the Fe $2p$ HAXPES and MCD-HAXPES spectra indicates that the peak shift originates from the hole doping by As ions. This result suggests that the doped As ions in bcc Fe act as acceptors.

The $\theta$ and magnetic field dependences of XMCD have been measured to investigate the MA of the bcc Fe-As film in detail [Figs. 6(a) and 6(b)]. The Fe $L_3$ XMCD intensity for high magnetic fields taken at $\theta = 0°$ is slightly higher than that taken at $\theta = 60°$. Since the XAS spectra are independent of the direction of the magnetic field within experimental accuracy, this difference would not originate from the saturation effect, which is a phenomenon that the intensity of the XAS peak decreases when the X-ray is subsequently absorbed with propagating the materials [35]. Thus, the observed $\theta$ dependence of the Fe $L_3$ XMCD reflects the inherent magnetic properties of the Fe-As film. The origin of this difference shall be discussed below. As for the As $L_{2,3}$ XMCD spectra, the spectral line shapes taken at $\theta = 0°$ are similar to those taken at $\theta = 60°$ within the experimental accuracy, as shown in Fig. 5(b). In order to analyze the magnetic characteristics of Fe and As, the $\theta$ dependence of the XMCD-$H$ curves has also been measured [see Figs. 6(a) and 6(b)]. The XMCD-$H$ curves of Fe (orange curve, left axis) are taken at the $L_3$ edge. The Fe $L_{2,3}$ XMCD-$H$ curve taken at $\theta = 60°$ under a low magnetic field more steeply increases than those taken at $\theta = 0°$, indicating that the ferromagnetic Fe component shows IMA as well as the results of SQUID. The XMCD-$H$ curves of As (blue curve, right axis) are obtained by subtracting the As $L_3$ XMCD from the As $L_2$ XMCD due to the small XMCD intensity; the difference of XMCD between the $L_2$ and $L_3$ edge makes the XMCD intensity effectively large since the sign of the $L_2$ XMCD is opposite to the $L_3$ XMCD. The hysteresis curves of Fe are qualitatively the same as those of As. It should be noted here that the finite coercivity of the As $L$ XMCD-$H$ curve is observed as well as that of the Fe $L$ XMCD-$H$ curve, as shown in the inset of Fig. 6(b), providing the experimental evidence that the doped As ions contribute to the ferromagnetism of Fe-As. Based on the findings, the magnetic moment of the As ions is considered to be induced via the orbital hybridization, i.e., the *p-d* hybridization in bcc Fe-As. Note that the XMCD-$H$ curves taken at $\theta = 0°$ and $60°$ slightly increase at the high magnetic field region, indicating that the Fe-As also contains a small amount of the paramagnetic Fe component, which is ignored in the SQUID results. In Fig. 6(c), the magnetic moment of the ferromagnetic ($m_{Ferro.}$) and paramagnetic ($m_{Para.}$) components at $\mu_0 H = 4$ T is estimated by fitting the linear paramagnetic component of the Fe $L_{2,3}$ XMCD-$H$ curve taken at $\theta = 0°$ in the region between $\mu_0 H = 2$ T and 4 T. The value of $\frac{m_{Ferro}}{m_{Ferro}+m_{Para}} = 0.971$ at $\mu_0 H = 4$ T indicates that the ferromagnetic component is dominant in the Fe-As film.

We analyze the Fe $L_{2,3}$ XAS and XMCD spectra to estimate the hole number of the Fe ions. One can calculate the orbital ($m_{orb,\theta}$) and effective spin ($m_{spin,\theta}^{eff}$) magnetic moments from XAS and XMCD spectra by applying the XMCD sum rules [36]:

$$m_{orb,\theta} = -\frac{2\int_{L_{2,3}}\Delta\mu_\theta d\omega}{3\int_{L_{2,3}}\bar{\mu}_\theta d\omega}n_h,$$ (1)

$$m_{spin,\theta}^{eff} = m_{spin} + \frac{7}{2}m_{T,\theta} = -r\frac{3\int_{L_3}\Delta\mu_\theta d\omega - 2\int_{L_{2,3}}\Delta\mu_\theta d\omega}{\int_{L_{2,3}}\bar{\mu}_\theta d\omega}n_h.$$ (2)

$m_{spin}$, $m_{T,\theta}$, $r$, and $n_h$ are the spin magnetic moment, the magnetic dipole term [37], the correction factor, and the number of the holes, respectively . In general, $\theta$ dependence of $m_{orb}$ and $m_T$ reflect the shape of the Fe $3d$ orbitals and the anisotropy of the spin-density distribution, respectively. On the other hand, $m_{spin}$ is independent of $\theta$ since the weak spin-orbit coupling in the $3d$ transition metals such as Fe leads to the practically isotropic spin magnetic moment [38]. First, we estimate the saturation magnetization of Fe ($m_{Fe} = m_{spin}+m_{orb}$) from the XMCD sum rule Eqs. (1) and (2). Due to the XMCD geometry, $m_{T,\theta}$ is given by the following equations [38–40]:

$$m_{T,\theta} = m_{T,\perp} \cos^2 \theta + m_{T,\parallel} \sin^2 \theta \,, \tag{3}$$

$$m_{T,\perp} + 2m_{T,\parallel} = 0, \tag{4}$$

where $\perp$ ( $\parallel$ ) represents $\theta = 0°$ (90°). The ferromagnetic Fe component in the Fe-As film has isotropic $m_{orb}$ since the XMCD integrals are independent of $\theta$ within the experimental accuracy [Fig. 4(a)].

$$m_{orb} = m_{orb,\perp} = m_{orb,\parallel} = m_{orb,60°} \tag{5}$$

By combining Eqs. (3) and (4) for $\theta = 60°$, we obtain

$$m_{T,60°} = -\frac{1}{8} m_{T,\perp}. \tag{3}'$$

Eq (3)' gives the following representation of Eq. (2):

$$m_{spin,60°}^{eff} = m_{spin} - \frac{7}{16} m_{T,\perp}, \tag{2}'$$

$$m_{spin,\perp}^{eff} = m_{spin} + \frac{7}{2} m_{T,\perp}. \tag{2}''$$

We can obtain $m_{spin}$ from Eqs. (2)' and (2)":

$$m_{spin} = \frac{m_{spin,\perp}^{eff} + 8m_{spin,60°}^{eff}}{9}. \tag{6}$$

Using Eqs. (5) and (6), $m_{Fe}$ is given by

$$m_{Fe} = \frac{m_{spin,\perp}^{eff} + 8m_{spin,60°}^{eff}}{9} + m_{orb,\perp}. \tag{7}$$

Based on Eqs. (1), (2), and (7), we can accurately estimate $m_{Fe}$ by eliminating the influence of $m_T$ from the XAS and XMCD spectra shown in Fig. 4(a). Second, we combine the saturation magnetization of the ferromagnetic component obtained by SQUID ($m_{sat,\text{SQUID}}$) and XMCD ($m_{sat,\text{XMCD}}$) to calculate the amount of changed $n_h$ by As doping ($\Delta n_h$). Since the ratio of Fe to As atoms is 3:1, $m_{sat,\text{XMCD}}$ is represented by

$$m_{sat,\text{XMCD}} = \frac{m_{Ferro}}{m_{Ferro}+m_{Para}}(0.75m_{\text{Fe}} + 0.25m_{\text{As}}) = \frac{m_{Ferro}}{m_{Ferro}+m_{Para}} \times 0.75m_{\text{Fe}}. \quad (8)$$

Here, $m_{\text{As}}$ is the saturation magnetization of As, and we assume $m_{\text{As}}$ is negligibly smaller than $m_{\text{Fe}}$, that is, $m_{\text{As}} \sim 0$ (see the magnitudes of the XMCD intensities shown in Fig. 4). Equation (8) includes the coefficient $\frac{m_{Ferro}}{m_{Ferro}+m_{Para}}$ since the magnetic moment estimated by the XMCD sum rules consists of the ferromagnetic and paramagnetic components. Based on Eq. (8) and $m_{sat,\text{SQUID}} \cong m_{sat,\text{XMCD}}$, we can obtain

$$m_{sat,\text{SQUID}} = \frac{m_{Ferro}}{m_{Ferro} + m_{Para}} \times 0.75m_{\text{Fe}}. \quad (9)$$

Here, we use the approximation that $n_h$ is represented by $\Delta n_h + 6.61$ ($n_h$ of Fe metal is 6.61 [31]) and $r$ is 1 ($r$ of Fe metal is 1 [31]). $\Delta n_h$ is estimated at $0.12 \pm 0.03$ from Eqs. (1), (2), (7) and (9). That is, the electronic configuration of Fe is changed from $3d^{3.39}$ to $3d^{3.27}$ by As doping, consistent with the MCD-HAXPES result that As is an acceptor in bcc Fe. The value of $n_h$ ($6.73 \pm 0.03$) gives $m_{\text{Fe}} = 2.37 \pm 0.02$ $\mu_{\text{B}}$/Fe. The estimated $m_{\text{Fe}}$, $m_{spin}$, $m_T$, and $m_{orb}$ are summarized in Table I. The analyses reveal that Fe-As has isotropic $m_{orb}$ and anisotropic $m_T$.

Based on the experimental findings, let us discuss the role of As ions in the MA of Fe-As. In the phenomenological model of the MA of a ferromagnet with cubic and uniaxial magnetocrystalline anisotropy components, the anisotropy energy ($E_\theta$) along the $\theta$ direction can be written as

$$E_\theta = K_1[(\alpha_1\alpha_2)^2 + (\alpha_2\alpha_3)^2 + (\alpha_3\alpha_1)^2] + K_2(\alpha_1\alpha_2\alpha_3)^2 - K_u(\boldsymbol{M}_s \cdot \boldsymbol{u}/M_s)^2. \quad (10)$$

Here, $K_1$ and $K_2$ are the cubic anisotropy constants, $K_u$ is the uniaxial anisotropy constant, the magnetization vector $\boldsymbol{M}_s$ is defined by the direction cosines $\alpha_1$, $\alpha_2$, and $\alpha_3$ with respect to the cubic axes. The unit vector $\boldsymbol{u}$ represents the direction of the uniaxial anisotropy direction. $E_\theta$ is decomposed into the cubic ($E_{C,\theta}$) and uniaxial ($E_{U,\theta}$) components. That is, $E_\theta = E_{C,\theta} + E_{U,\theta}$, where

$$E_{C,\theta} = K_1[(\alpha_1\alpha_2)^2 + (\alpha_2\alpha_3)^2 + (\alpha_3\alpha_1)^2] + K_2(\alpha_1\alpha_2\alpha_3)^2, \qquad (11)$$

$$E_{U,\theta} = K_u(\boldsymbol{M}_s \cdot \boldsymbol{u}/M_s)^2. \qquad (12)$$

We define the differences of the cubic magnetic anisotropy energy (CMA, $E_{CMA}$) and uniaxial magnetic anisotropy energy (UMA, $E_{UMA}$) as $E_{C,\perp} - E_{C,\parallel}$ and $E_{U,\perp} - E_{U,\parallel}$, respectively. Here, the $\perp$ and $\parallel$ directions are parallel to the [111] and [11$\bar{2}$] axis. In order to estimate the order of magnitude of $E_{CMA}$ of the Fe-As film, we use $E_{CMA}$ of bcc Fe metal. $E_{CMA}$ of bcc Fe metal is estimated as $4.2 \times 10^{-3}$ [MJ/m$^3$] from Eq. (11), where $K_1$ is $5.0 \times 10^{-2}$ [MJ/m$^3$] and $K_2$ is quite smaller than $K_1$ [41]. The order of magnitude of $E_{CMA}$ of the Fe-As film is assumed as $10^{-3}$ [MJ/m$^3$]. In order to estimate the $E_{UMA}$, we adopt the extended Bruno model [42] since $\Delta m_{orb}$ ($= m_{orb,\perp} - m_{orb,\parallel}$) is nearly zero and $\Delta m_T$ ($= m_{T,\perp} - m_{T,\parallel}$) is a finite value. The extended Bruno model is as follows:

$$E_{UMA} = \frac{\xi}{4\mu_\text{B}} \Delta m_{orb} - \frac{21}{2\mu_\text{B}} \frac{\xi^2}{\Delta_{ex}} \Delta m_T, \qquad (10)$$

where $\xi$ and $\Delta_{ex}$ are the spin-orbit coupling constant and the exchange splitting of 3$d$ bands, respectively [43]. The first and second terms represent the anisotropy of the orbital magnetic moment and the spin-density distribution, respectively. Since the value of $\Delta m_{orb}$ estimated by the XMCD sum rules is nearly zero, the first term in Eq. (10) has less contribution to the magnetic anisotropy and the second term primarily determines the MA in Fe-As. The estimated value of $E_{UMA}$ is $-5.5 \pm 0.5$ [MJ/m$^3$] using the values of $\xi =$ 0.05 eV of Fe$^{3+}$ ions [44,45] and $\Delta_{ex} = 1.5$ eV of metal Fe [46,47]. The negative value of the $E_{UMA}$ indicates that Fe-As has IMA. Note that IMA of Fe-As mainly originates from the intrinsic UMA rather than the CMA and the shape anisotropy for the thin film, because the estimated $E_{UMA}$ is several times larger than the energy of the shape anisotropy $E_{SA}$ ($= -1.3 \pm 0.1$ [MJ/m$^3$]) and substantially larger than $E_{CMA}$. The $m_T$-driven MA of Fe-As is different from the spin-orbit interaction and/or orbital-magnetic-moment-driven MA of other Fe-based materials, such as FePt alloy [24,25] and Fe ultrathin films [14,26], and the lattice-expansion model in Fe$_x$N [11,27]. It should be mentioned here that $m_T$ of the bcc Fe metal is approximately zero [38]. The possible origins of the anisotropy of $m_T$ of Fe-As can be attributed to the epitaxial strain and the $p$-$d$ hybridization. The anisotropy of $m_{orb}$ of Fe ion is smaller than the experimental accuracy of the XMCD measurements, that is, 0.01 $\mu_B$/Fe. This can be explained by the following reason: metallic bonds dominate Fe ions due to more Fe-Fe bonds than Fe-As bonds, and the crystal field is more effective for ionic bonds than metallic bonds. On the other hand, the epitaxial strain modulates the electronic structure of As ions by the change of the crystal field since As ions are surrounded by Fe ions and ionic bonds are dominant. Due to the compressive strain from the substrate, the As $4p$ orbitals split into $4p_z$ ($z$ // [111]) and two-fold $4p_x$/$4p_y$ orbitals due to the change of the crystal field from the cubic to tetragonal symmetry, and the $4p_z$ orbitals are more stable than the two-fold $4p_x$/$4p_y$ orbitals. The orbital splitting and hole doping to Fe ions make the As $4p$ configuration of $4p_x^1 p_y^1 p_z^{1.6}$, leading to the anisotropic As $4p$ distribution along the out-of-plane axis [see Figs. 7(a) and 7(b)]. Moreover, the compressive strain elongates the bonding length of Fe and As ions along the [111] axis and shortens the distance between Fe and As ions in the (111) plane [see Fig. 7 (a)]. This difference makes the larger orbital overlap of Fe $3d$ and As $4p_x$/$4p_y$ orbitals than that of Fe $3d$ and As $4p_z$ orbitals, leading to the stronger $p_x$/$p_y$-$d$ hybridization than the $p_z$-$d$ one. Considering that the As $4p$ orbital is hybridized with the minority spin state of the Fe $3d$, the spin direction of $4p_x$/$p_y$ electrons is opposite to the majority spin state of the Fe $3d$, which is consistent with the XMCD results. Via the $p$-$d$ hybridization, the anisotropic As $4p$ distribution probably causes the anisotropic spin distribution of Fe $3d$ electrons along the out-of-plane [111] axis, resulting in the anisotropic $m_T$. Our findings suggest that the orbital hybridization between Fe and doped non-magnetic ions can control the MA of the bcc Fe thin film. This mechanism of the MA is based not on the unique properties of Fe and As ions but on the orbital hybridization of magnetic $3d$ transition metal and non-magnetic ions and epitaxial strain. That is, the suggested mechanism is possibly applied to other magnetic $3d$ transition metal thin films doped with non-magnetic ions.

## 4. Summary

We have unveiled the role of As ions in the MA of the Fe-As thin film grown on an SI-GaAs(111)B substrate by XMCD and MCD-HAXPES. The MCD-HAXPES spectra of the Fe $2p$ core level suggest that the As ions act as acceptors in the Fe-As film. The XMCD spectra at the Fe $L_{2,3}$ edge and the As $L_{2,3}$ edge demonstrate that the As ions contribute to the ferromagnetism through hybridization between the Fe $3d$ and As $4p$ orbitals. We estimate the isotropic orbital magnetic moment and the anisotropic magnetic dipole term of Fe by the XMCD sum rules. and reveal that the anisotropic magnetic dipole term contributes to the IMA of Fe-As by evaluating the cubic, uniaxial, and shape magnetic anisotropy energy. The anisotropy of the magnetic dipole term can be attributed to the anisotropic $p$-$d$ hybridization due to epitaxial strain. Our results suggest that the MA of the bcc Fe thin film can be controlled through orbital hybridization between Fe and non-magnetic elements and epitaxial strain, being possibly suitable for other magnetic $3d$ transition metal thin films doped with non-magnetic elements.

## ACKNOWLEWDGMENTS

This work was supported by Grants-in-Aid for Scientific Research (No. 19K21961, 20H05650, 22K03535, 23K17324, 24H00018) and by JST-CREST (JPMJCR1777). This work was partially supported by the Spintronics Research Network of Japan (Spin-RNJ). This work was performed under the Shared Use Program of Japan Atomic Energy Agency (JAEA) Facilities (Proposal No. 2021A-E24) supported by JAEA Advanced Characterization Nanotechnology Platform as a program of "Nanotechnology Platform" of the Ministry of Education, Culture, Sports, Science and Technology (MEXT) (Proposal No. JPMXP09A21AE0022). Supporting experiments at SPring-8 were approved by the Japan Synchrotron Radiation Research Institute (JASRI) Proposal Review Committee (Proposal No. 2021A3841, 2022B1761). A.F. acknowledges support from National Science and Technology Council of Taiwan (NSTC 113-2112-M-007-033).

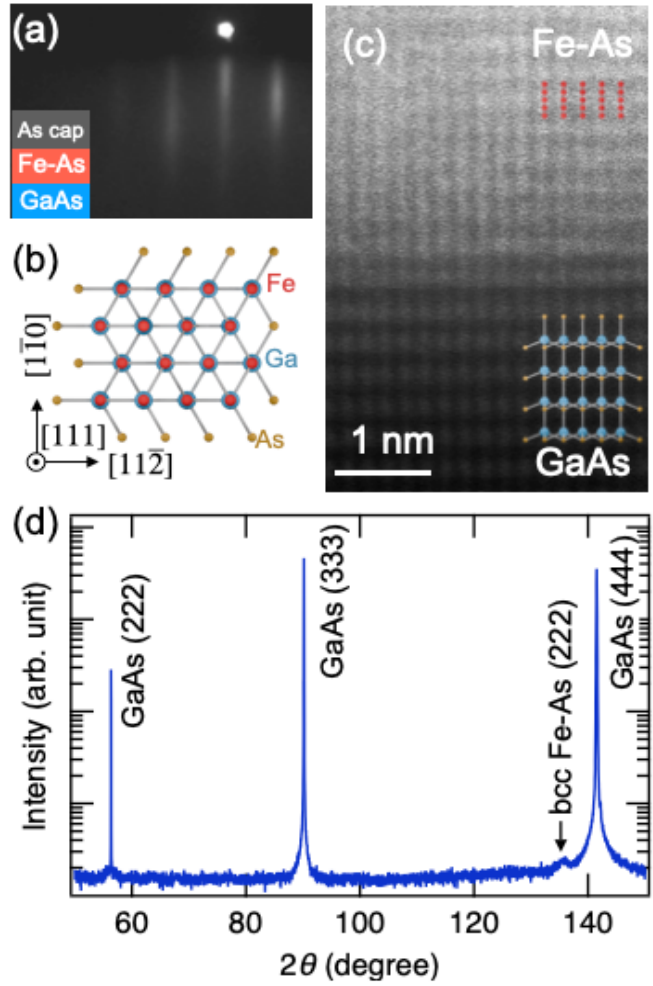

FIG. 1: Crystal structure of the Fe-As film. (a) RHEED pattern of the Fe-As film in the $[\bar{1}10]$ direction. Inset shows the heterostructure of As-cap/Fe-As/SI GaAs substrate. (b) Epitaxial relation of bcc Fe-As and GaAs visualized by VESTA [48]. (c) HAADF-STEM image of the Fe-As film on the GaAs substrate projected along the $[11\bar{2}]$ axis. (d) XRD $2\theta$-$\omega$ scan of the Fe-As film. The arrow denotes the diffraction peak from bcc Fe-As (222).

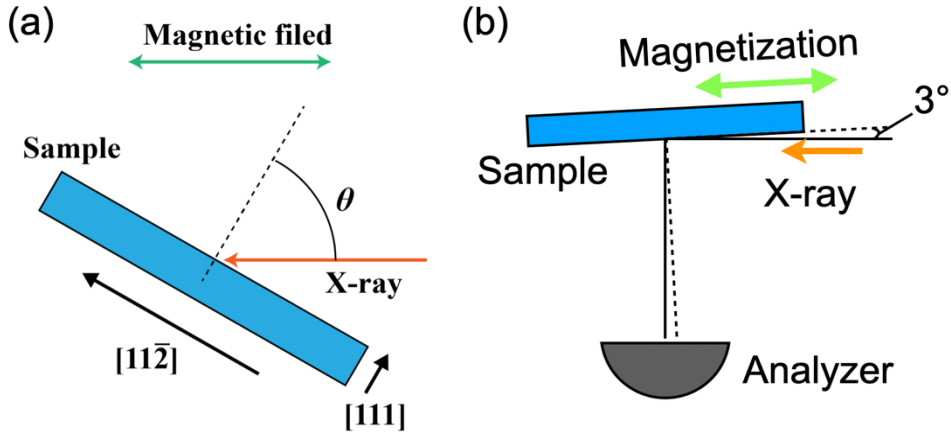

FIG. 2: Experimental geometry. Setup of (a) XMCD and (b) MCD-HAXPES.

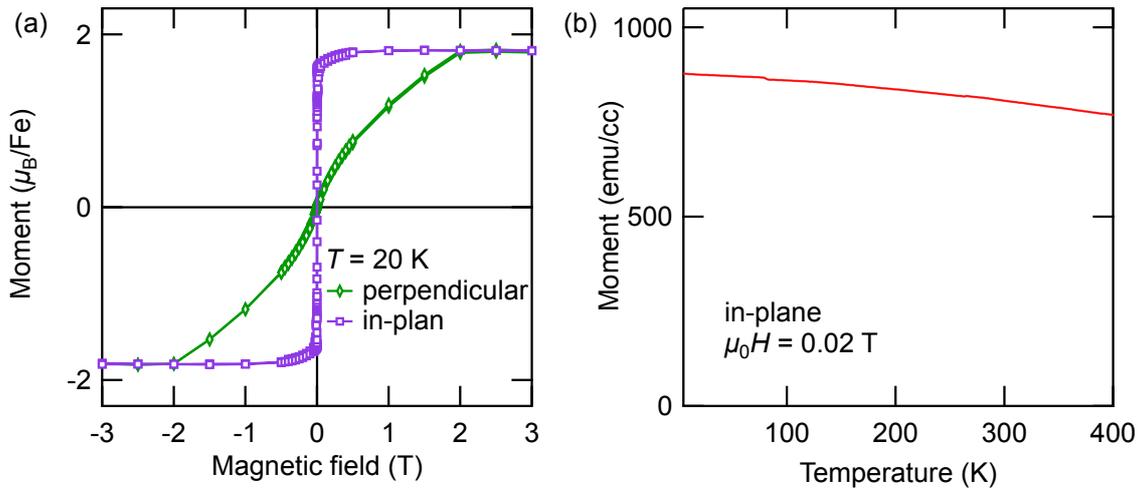

FIG. 3: Magnetic properties of the Fe-As film. (a) Hysteresis curves measured by SQUID, where the directions of the magnetic field are perpendicular to the film plane (green curve) and in the film plane (purple curve). Magnetic moments per Fe atom are shown/plotted. (b) $m$-$T$ curve at $\mu_0 H = 0.02$ T.

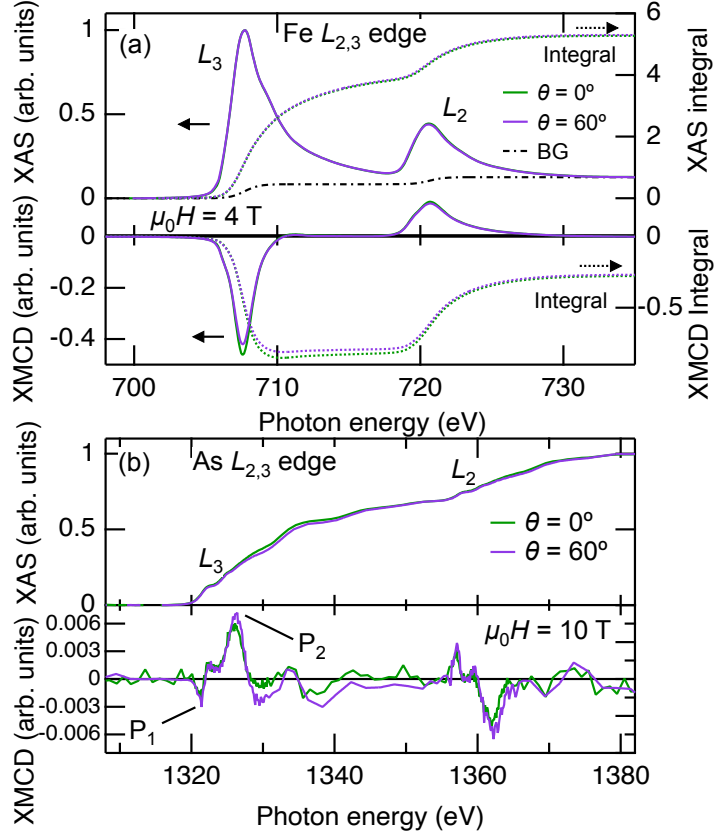

FIG. 4: XAS and XMCD spectra of Fe and As. Angle dependence of XAS and XMCD spectra at the (a) Fe and (b) As $L_{2,3}$ absorption edge under $T = 20$ K, where green and purple curves represent the spectra taken at $\theta = 0°$ and $\theta = 60°$, respectively. The XAS and XMCD spectra of Fe and As are taken at $\mu_0 H = 4$ T and 10 T, respectively. The dashed and dash-dotted curves mean the integral of XAS (XMCD) and background (BG).

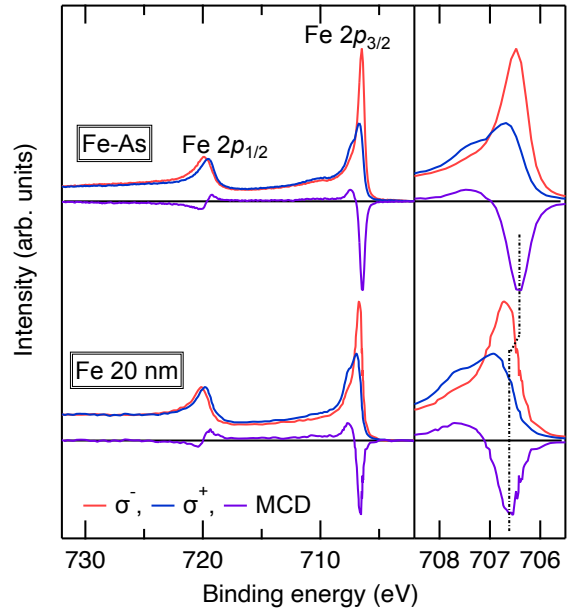

FIG. 5: HAXPES and MCD-HAXPES spectra of Fe 2$p$ of Fe-As (upper), 20 nm Fe film [34] (lower), where blue and red curves represent the Fe 2$p$ spectra taken by $\sigma^+$ and $\sigma^-$-polarized X-rays, respectively and black curves are the MCD spectra (left panel). In the right panel, the Fe 2$p_{3/2}$ spectra are enlarged.

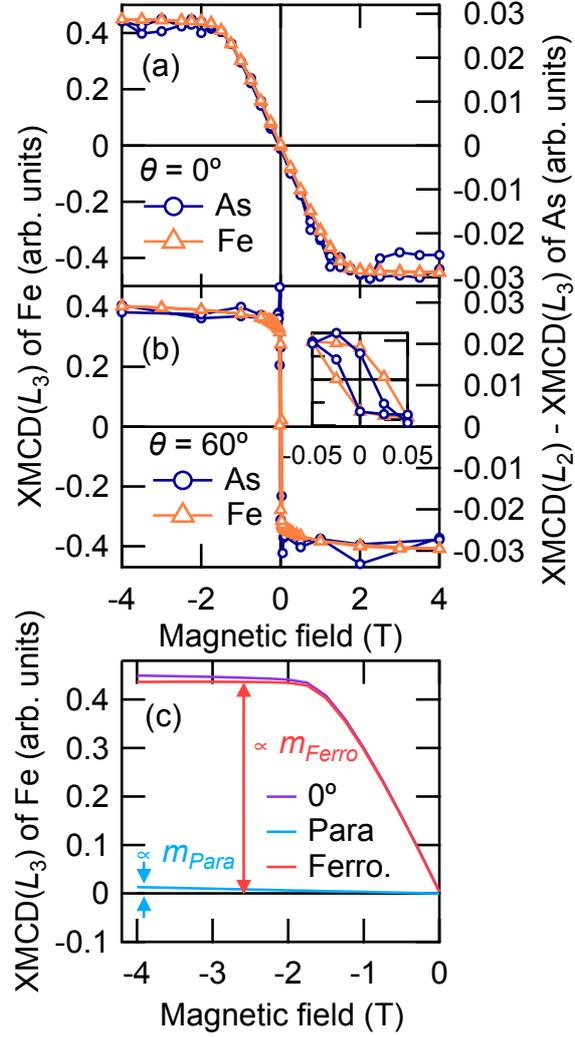

FIG. 6: Magnetic field dependence of XMCD of Fe and As. XMCD-$H$ curves taken at (a) $\theta = 0°$ and (b) $\theta = 60°$. Here, the XMCD-$H$ curve of Fe (orange curve, left axis) is taken at the $L_3$ edge, and the XMCD-$H$ curves of As are represented by subtracting the XMDC-$H$ curve at the $L_3$ edge from that at the $L_2$ edge. The inset of (b) shows the enlarged XMCD-$H$ curve around $\mu_0 H = 0$ T. (c) Decomposition of the XMCD-$H$ curve at the Fe $L_3$ edge taken at $\theta = 0°$ (purple curve) into the ferromagnetic and paramagnetic components, where light-blue and red curves represent the paramagnetic and ferromagnetic components.

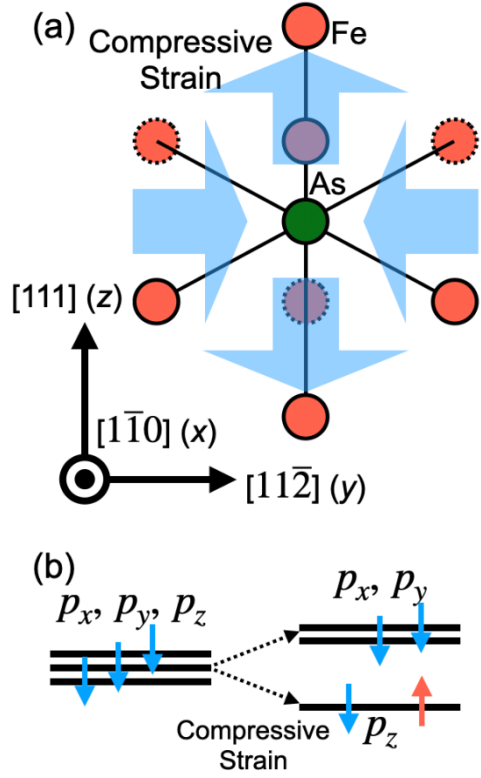

FIG. 7: Effect of strain on Fe-As. (a) Distortion of crystal structure. Here, red and green circles represent the Fe and As ions, respectively, and the dashed circles mean that the ions are located at the back side of the paper. Blue arrows show the direction of the distortion due to the compressive strain. (b) Energy splitting of the As 4$p$ orbitals by the crystal field of the compressive strain. Blue and red arrows mean minority (down) and majority (up) spins.

TABLE I. Spin and orbital magnetic moments of Fe in Fe-As. All the values have been estimated using the XMCD sum rules [Eqs. (1) and (2)].

| | $m_{orb}$ [$\mu_\mathrm{B}$/Fe] | $m_{spin}$ [$\mu_\mathrm{B}$/Fe] | $m_T$ [$\mu_\mathrm{B}$/Fe] | $m_\mathrm{Fe}$ [$\mu_\mathrm{B}$/Fe] |
|---|---|---|---|---|
| $\theta = 0°$ ($\perp$) | $0.22 \pm 0.01$ | $2.15 \pm 0.01$ | $0.02 \pm 0.01$ | $2.37 \pm 0.02$ |
| $\theta = 90°$ ($\parallel$) | $0.22 \pm 0.02$ | $2.15 \pm 0.01$ | $-0.01 \pm 0.01$ | $2.37 \pm 0.02$ |